\documentclass[10pt,twoside]{article}
\usepackage{pkas2}
\usepackage{graphicx}
\makehead{Serjeant}{A multi-wavelength view of galaxy evolution with AKARI}

\def\lesssim{\mathrel{\hbox{\rlap{\hbox{\lower4pt\hbox{$\sim$}}}\hbox{$<$}}}}
\def\gtrsim{\mathrel{\hbox{\rlap{\hbox{\lower4pt\hbox{$\sim$}}}\hbox{$>$}}}}

\begin{document}
\sloppy
\pagenumbering{arabic}
\twocolumn[
\pkastitle{27}{1}{2}{2012}
\begin{center}
{\large \bf {\sf
A MULTI-WAVELENGTH VIEW OF GALAXY EVOLUTION WITH AKARI}}\vskip 0.5cm
{\sc S.Serjeant$^{1}$, C.Pearson$^{1,2}$, G.J.White$^{1,2}$,
  M.W.L.Smith$^3$, 
Y.Doi$^4$
}\\
$^1$Department of Physical Sciences, The Open University, Milton Keynes, MK7 6AA, UK\\
$^2$RAL Space, STFC Rutherford Appleton Laboratory, Chilton, Didcot,
Oxfordshire, OX11 0QX, UK\\
$^3$School of Physics and Astronomy, Cardiff University, Queens Buildings,
The Parade, Cardiff CF24 3AA, UK\\
$^4$Department of General System Studies, Graduate School of Arts and Sciences,
The University of Tokyo, 3-8-1 Komaba, Meguro-ku, Tokyo 153-8902, Japan\\
\normalsize{\it (Received July 17, 2012; Accepted ????)}
\end{center}
\newabstract{ AKARI's all-sky survey resolves the far-infrared
  emission in many thousands of nearby galaxies, providing essential
  local benchmarks against which the evolution of high-redshift
  populations can be measured. This review presents some recent
  results in the resolved galaxy populations, covering some well-known
  nearby targets, as well as samples from major legacy surveys such as
  the Herschel Reference Survey and the JCMT Nearby Galaxies
  Survey. This review also discusses the prospects for higher
  redshifts surveys, including strong gravitational lens clusters and
  the AKARI NEP field.  \vskip 0.5cm {\em key words:} infrared -
  telescope: conferences - proceedings} \vskip 0.15cm \flushbottom ]

\newsection{INTRODUCTION} 

This paper will review results on bright galaxies from AKARI and
Herschel, and will cover from local galaxies to high-z galaxies, via
gravitational lensing. 

The first part of the review will be about the AKARI XTENDED
Prioritised Study (PS) program. We have heard a great deal about the
AKARI diffuse maps so far in this conference and this paper will focus
on some early results on resolved galaxies. We will cover a few
obvious old friends, such as Andromeda, and then progress from these
anecdotal examples to discussing results from two major legacy surveys
with robust selection criteria. Note that no claim will be made that
any of these surveys are unbiased in any sense. There is no such thing
as an unbiased survey in astronomy. A `bias' is just a pejorative way
of referring to the selection effects, and any astronomical catalogue
of any nature has selection criteria of some sort.  (Even ``all
luminous objects within this volume'' would neglect neutral hydrogen
systems, and would in practice have a luminosity limit anyway;
astronomical surveys are defined by what they exclude, not what they
include.)  Instead of trying to perform the impossible feat of
avoiding `biases', one must understand and quantify one's selection
effects, which is the key advantage of these surveys over heterogenous
compilations.

Having described some of the early results from the XTENDED PS
program, the focus will move onto the Herschel ATLAS survey. This is
in some ways a complementary project to the AKARI all-sky survey, and
has mapped 1\% of the sky to almost as shallow a depth as Herschel can
achieve.  Herschel ATLAS covers local as well as high
redshift galaxies, and we'll go from low redshift to high redshift via
gravitational lensing. Staying on the lensing theme, this review will
discuss 
recent evidence for the links between the Herschel
and AKARI populations that have been found using the cluster lens
Abell 2218. Finally, this paper reviews the prospects for Herschel
data in the AKARI NEP-Deep field, in the light of the current results
from Herschel ATLAS and other surveys.

%
%

\newsection{THE AKARI XTENDED PRIORITISED STUDY} 

\begin{figure}[!ht]
\resizebox{\hsize}{!}{\includegraphics{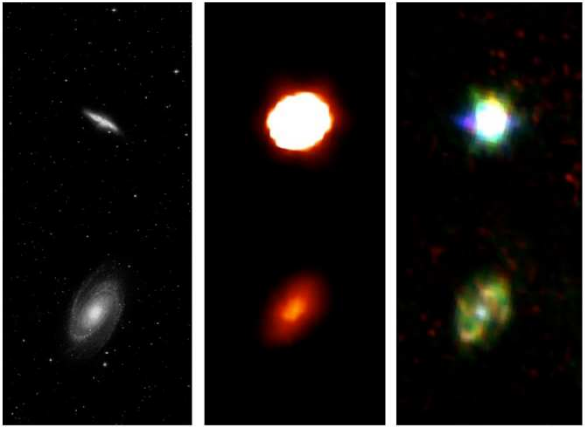}}
\caption{\label{fig:m82}
M82 (top) and M81 (bottom) in DSS B-band (left), IRAS $60\,\mu$m
(centre) and in the
  AKARI all-sky survey (right, with $160,140,90\,\mu$m as RGB
  respectively). Note the arms in M81 resolved by AKARI. The two
  galaxies are separated by about $37'$, or $(38\pm5)\,$kpc in
  projection.} 
\end{figure}

\begin{figure}[!ht]
\resizebox{\hsize}{!}{\includegraphics{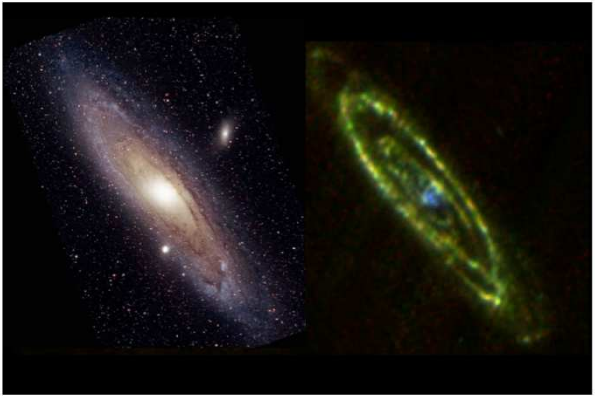}}
\caption{\label{fig:m31} M31 in the optical (left; NASA APOD, Jason
  Ware) and in the AKARI all-sky survey (right, with
  $160,140,90\,\mu$m as RGB respectively). The diameter of M31 is
  approximately $30$\,kpc, and $1^\circ\equiv13.6\,$kpc. Note the
  star-forming ring $\sim10$\,kpc from the centre, and the warm
  central dust (also associated with a CO deficit).}
\end{figure}

The goal of the XTENDED program is the scientific exploitation of an
all-sky far-infrared atlas of resolved galaxies from AKARI. Unlike
e.g. the heterogenous SINGS survey in which the selection criteria are
difficult to quantify (Kennicutt et al., 2003), the sample
is limited purely by far-infrared surface brightness. The project will
test the frequency of occurence of cool extended dust components
undetectable by IRAS, construct spatially-resolved radio/far-infrared
correlations, map every known blue compact dwarf galaxy, and help
determine the extent to which integrated properties of galaxies are
reliable measures of the mean physical conditions within them.  Work
on the whole sample is still ongoing so this paper will discuss only a
few key targets.  There are rich seams to be
mined. Figure \ref{fig:m82} compares AKARI diffuse maps and IRAS images
of the prototypical starburst M82, plus its companion M81. What is
immediately clear is the improved angular resolution of AKARI, with
the arms of M81 clearly discernable. AKARI succeeds in resolving the
dust emission in many thousands of local galaxies.

Figure \ref{fig:m31} shows another old friend, M31, at optical and
far-infrared wavelengths. The morphology agrees with ISO and
Spitzer far-infrared observations (Haas et al., 1998, Gordon et
al., 2006).  M31 has the interesting property that $90\%$ of the 
far-infrared emission is not associated with star formation.  There is
a higher temperature knot in the cold dust distribution at the centre
of M31, associated with a deficit in CO, and there is also a
star-forming ring about 10\,kpc from the centre. The diameter of M31
is about 30\,kpc in this image.

\begin{figure}[!ht]
\resizebox{\hsize}{!}{\includegraphics{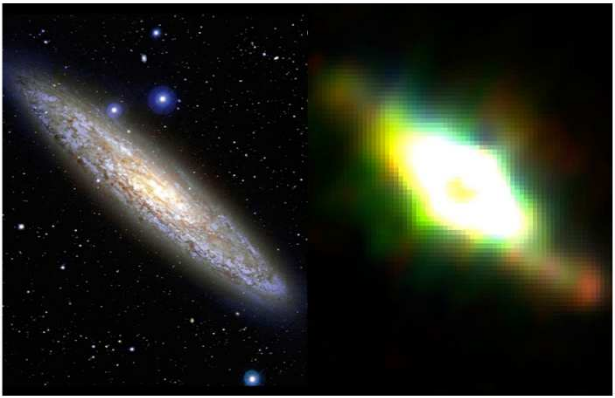}}
\caption{\label{fig:ngc253} 
Optical image of NGC\,253 (left; NASA APOD, CFHT) compared to the 
AKARI all-sky survey image (right) with $160,140,90\,\mu$m
rendered as RGB respectively. The optical diameter is
$27.5'\equiv(28\pm2)\,$kpc. Note  
the superwind and the central saturated pixel. 
}
\end{figure}

Figure \ref{fig:ngc253} presents optical and far-infrared data on
NGC\,253, which has far-infrared emission from a superwind driven by its
starburst. However note the yellow pixel in the centre, caused by
saturation; clearly, care needs to be taken with far-infrared
photometry of bright galaxies in the AKARI all-sky survey.

As a sanity check of the diffuse map fluxes, large aperture photometry
measurements ($4.75'$ radius) were taken
of Arp\,220 in the all-sky diffuse maps.
The flux measurements were $72$, $74$, $51$, $68$\,Jy (all $\pm1$\,Jy)
in the filters N60, WIDE-S, WIDE-L and N160 respectively, i.e. at
$65\,\mu$m, $90\,\mu$m, $140\,\mu$m and $160\,\mu$m
respectively. These are up to a factor of two fainter than catalogued
fluxes from ISO and IRAS, which we believe is due to detector
saturation in AKARI and/or incorrect flagging of peak fluxes as
glitches (see e.g. Yamamura et al., 2010), although no saturation
features such as that in Figure \ref{fig:ngc253} were obviously visible in the {\it images} in any
band. As a result we will restrict our discussion to targets with
predicted fluxes $<50$\,Jy per beam.



\newsection{AKARI OBSERVATIONS OF THE HERSCHEL REFERENCE SURVEY} 

So far we have covered only a few anecdotal examples of well-known
galaxies. There is only so far one can go with anecdotal evidence, so
the next stage is move to large samples with well-understood selection
criteria.  The first sample we will discuss is the Herschel Reference
Survey (Boselli et al., 2010). This is a guaranteed time programme on
Herschel to map about 300 local galaxies. It is volume limited and
K-band limited, implying an effective minimum stellar mass selection
depending on distance.  It has many science goals, including a census
of dust along the Hubble sequence, the connections between star
formation and dust emission, the global extinction in galaxies as a
function of type, the presence of dust in ellipticals and the cycle of
dust destruction and creation.  Ciesla et al., (2012) present submm
data for galaxies in this sample. The survey only obtained 
submm Herschel data with the SPIRE instrument, and as a result over a
third of their targets lack far-infrared data entirely. The AKARI
all-sky survey is ideal to address this need.

As part of the XTENDED PS program, fluxes for the Herschel Reference
Survey were extracted in a $300''\times300''$ box around each
galaxy. Sky subtraction was achieved by iteratively estimating the
mode of the pixel flux histogram in the $(1.5^\circ)^2$
postage stamps around each target, rejecting $>3\sigma$ or $<-3\sigma$
outliers. Flux errors were estimated by convolving the postage
stamp with a kernel equivalent to the photometric aperture, then
iteratively measuring the variance of the pixel count histogram of the
smoothed image.

\begin{figure}[!ht]
\resizebox{\hsize}{!}{\includegraphics{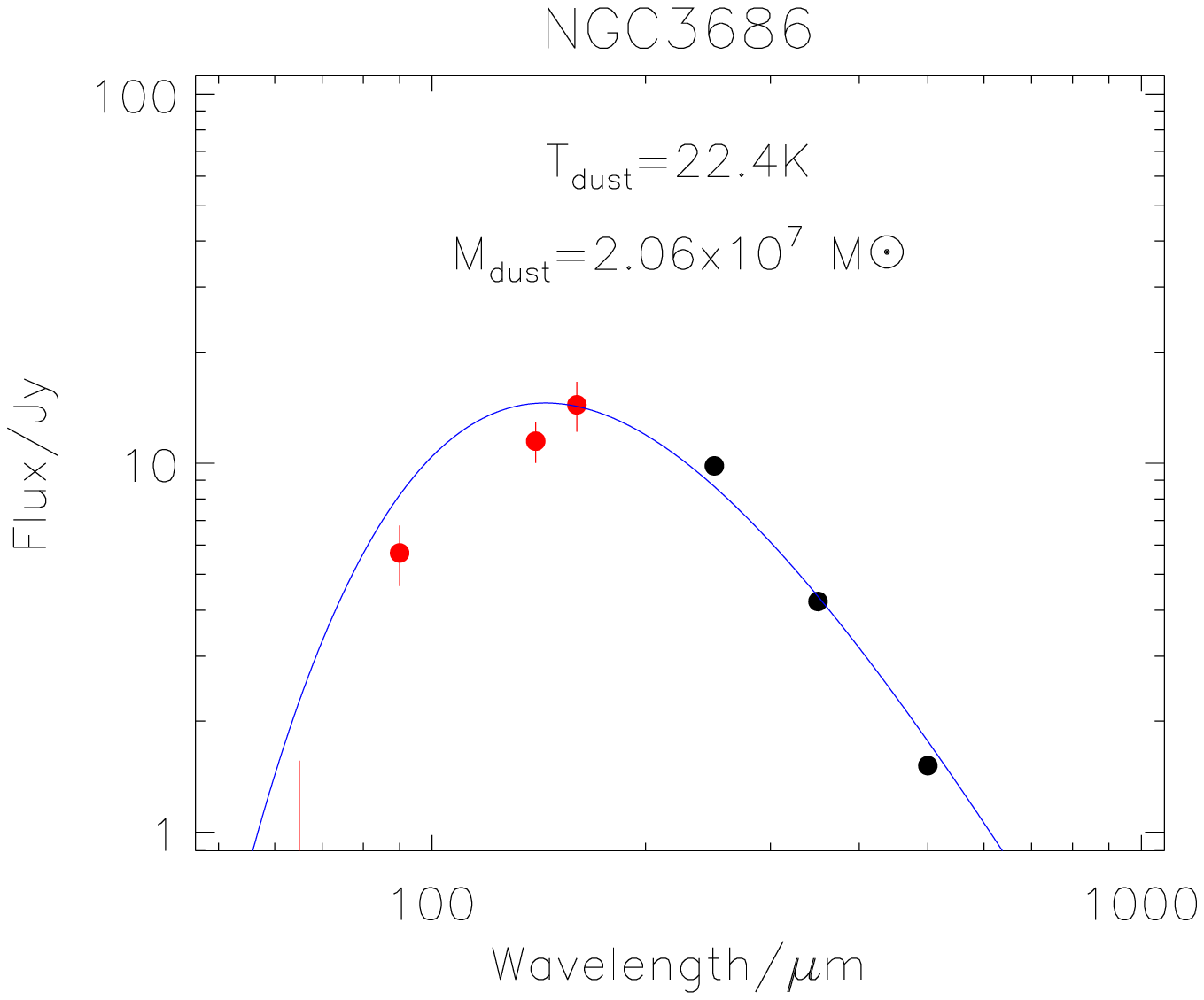}}
\resizebox{\hsize}{!}{\includegraphics{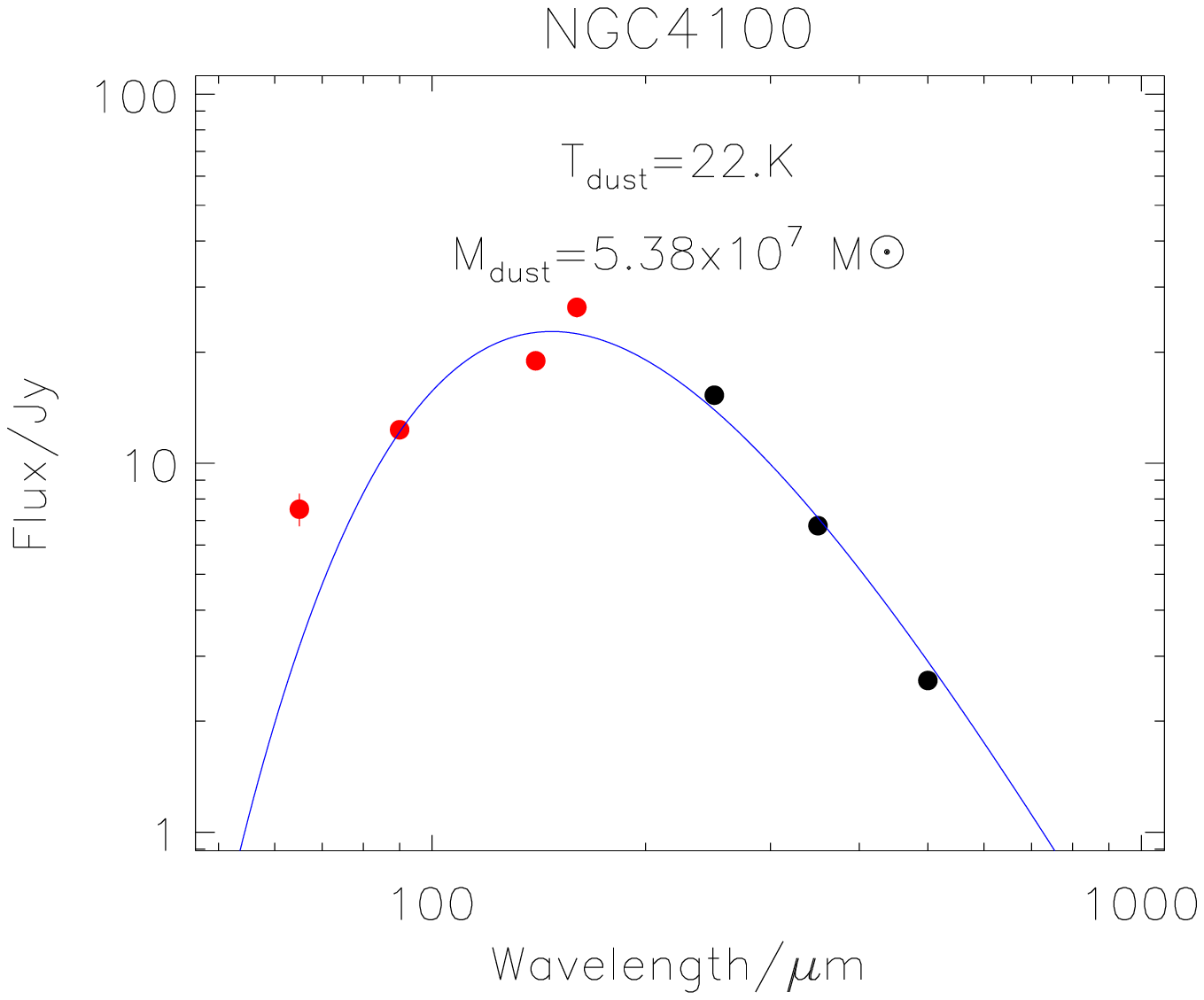}}
\caption{\label{fig:hrs} 
SEDs of two galaxies in the Herschel Reference Survey. The
$>200\,\mu$m data is from Herschel SPIRE, while $<200\,\mu$m is from
AKARI. The blue line is a grey body fit assuming an emissivity of
$\beta=1.5$. Note that the AKARI data clearly still has 
sources of systematic error of the order of $\sim30\%$ of unknown
origin, but possibly related to sky subtraction.}
\end{figure}

Careful sky subtraction was found to be important for the photometry,
but there are still unresolved problems. Figure \ref{fig:hrs} shows two
example SEDs from the combined Herschel and AKARI data from the
Herschel Reference Survey.
Clearly at least in the case of NGC\,4100 (and in other targets
not shown), there are still unresolved systematics in the fluxes; the
discrepancies with the models suggest the systematics are no more than
$\sim30\%$.  With this caveat in mind, we estimated dust masses
assuming single temperature grey-body fits (see e.g. Dunne et
al., 2011). Clearly, galaxies do not have single temperatures, or even
a few discrete temperatures; this is very much a ``spherical cow''
approximation. In some cases, an excess over the single-temperature
fits at shorter AKARI wavelengths requires the existence of warmer
dust phases. Future work will make use of radiative transfer
modelling. For the present purposes, the fits are used only to provide
order of magnitude estimates for dust masses, which will in any case
be dominated by the cooler components. The grey body fitting assumed
an emissivity of $\beta=1.5$. Dust mass estimates are typically a few
$\times10^7M_\odot$ in this sample. Work is ongoing to improve the
photometry in this sample.

\newsection{AKARI OBSERVATIONS OF THE JCMT NEARBY GALAXIES SURVEY} 

The JCMT Nearby Galaxies Survey (Wilson et al., 2012) is another major
legacy survey, with the goal of resolving sub-kpc structure in over
150 nearby galaxies. The goals, amongst others, are to see how
morphology and environment affect star formation in nearby
galaxies. The survey is covering the Spitzer SINGS sample (Kennicutt
et al., 2003), in order to benefit from the multi-wavelength legacy
data in SINGS. However, SINGS has a somewhat heterogeneous selection,
so the JCMT survey is also observing neutral-hydrogen-selected samples
in the field and in Virgo.  Again, the trouble with robust statistical
selection is that one is often left with not much supplementary data. In
particular, the field sample lacks far-infrared data. As with the
Herschel Reference Survey, the AKARI diffuse maps can provide this
far-infared data, sampling the peaks of the bolometric outputs of the
galaxies.

\begin{figure}[!ht]
\resizebox{\hsize}{!}{\includegraphics{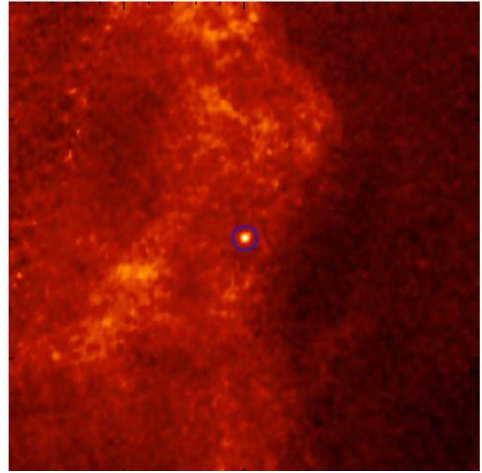}}
\caption{\label{fig:ngc1156} AKARI all-sky $90\,\mu$m image of
  NGC\,1156, from the JCMT Nearby Galaxies Survey. Note the
  surrounding cirrus structure. Lower angular resolution data would
  have difficulty decoupling the foreground cirrus and background
  galaxy. The image is $1.5^\circ$ on a side.}
\end{figure}

Figure \ref{fig:ngc1156} shows an example target from this survey, seen in
the AKARI all-sky survey. Note that AKARI's angular resolution has the
important advantage over IRAS that it can decouple flux from the
galaxy from foreground cirrus structure.

\begin{figure}[!ht]
\hspace*{-1cm}\resizebox{1.3\hsize}{!}{\includegraphics{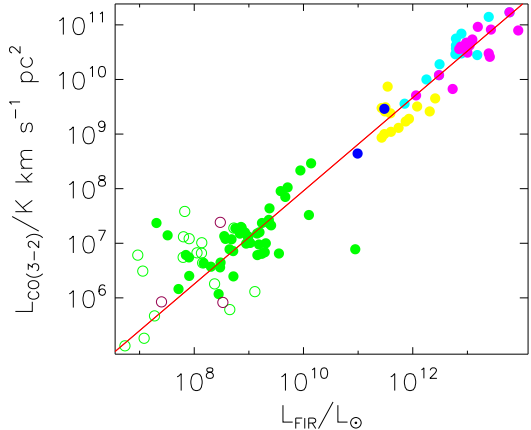}}
\caption{\label{fig:iono} CO(3-2) {\it vs.} far-infrared
  luminosity-luminosity correlation for the homogenously-selected Iono
  et al., (2009) data (luminous infrared galaxies: yellow,
  submm-selected galaxies: cyan, quasars: magenta, Lyman-break
  galaxies: blue), plus the AKARI $90\,\mu$m data on the JCMT nearby
  galaxies (spirals: green filled circles, irregulars: green open
  circles, others: brown open circles). The red power-law line is not
  quite a linear relation, and is an extrapolation from the Iono
  sample, not a fit to the JCMT data. Photometric errors in the JCMT
  sample are not shown for clarity but are consistent with the
  scatter.}
\end{figure}

Fluxes at $90\,\mu$m (WIDE-S filter) were extracted for all
the JCMT nearby galaxies field sample following the same procedure as
for the Herschel Reference Survey above, 
safely encompassing the optical diameters of the galaxies. 
Figure \ref{fig:iono} shows the CO(3-2) luminosities compared to the
far-infrared luminosities of homogenously-selected high-redshift
galaxies from Iono et al., (2009), plus the AKARI measurements of the
JCMT sample. The line is the extrapolation from the high-redshift
populations, and is not a fit to the JCMT data. This extends the Iono
et al., relation over five orders of magnitude in homogenously-selected
samples. The overall relation is not quite linear, implying that the
JCMT sample have far-IR to CO ratios lower than the ULIRGs in
general. One can convert the far-infrared to CO luminosity ratio into
a star formation rate to molecular gas mass ratio, resulting in a
timescale. From this, one can derive a molecular gas depletion
timescale of about 3\,Gyr in the JCMT sample. The gas depletion
timescale in submm galaxies is more like 0.5\,Gyr, once one has
remembered that different CO to H$_2$ conversions are appropriate for
ultraluminous infrared galaxies and submm galaxies.

\newsection{LOCAL GALAXIES AND LENSES IN HERSCHEL ATLAS}

The Herschel Astrophysical Terahertz Large Area Survey (H-ATLAS, Eales
et al., 2010) is perhaps Herschel's answer to the AKARI all-sky
survey. It has mapped about $1\%$ of the sky to almost the submm
confusion limits, discovering NGC4725 to have an ``Andromeda
analogue'' dust ring. Many hundreds of thousands of local
galaxies are expected in the survey; see e.g. Baes et al., 2010 for an
early example of a nearby galaxy with components of highly obscured
star formation which nonetheless contribute little to the global
extinction in the galaxy. AKARI detections of Herschel ATLAS local
galaxies are discussed at this conference in Pearson et al., (this
volume).  In roughly equal numbers at $500\,\mu$m to local galaxies,
H-ATLAS also made the landmark discovery of a large population of
strong gravitational lenses (Negrello et al., 2010), ultimately caused
by the steep intrinsic $500\,\mu$m bright number counts of
high-redshift galaxies. There are many examples of submm galaxies with
$z>2$ far-infrared photometric redshifts but identifications with
foreground spirals or ellipticals; not every obvious local galaxy
optical ID is the site of the observed far-infrared emission. HST,
IRAM, Herschel, SMA and other follow-ups are all ongoing.

\newsection{AKARI DEEP FIELDS: LINKING THE LOCAL AND HIGH-REDSHIFT
  UNIVERSE} 

A further strong gravitational lensing system, studied by both AKARI
and Herschel, is the galaxy cluster Abell\,2218. At the 2009 AKARI
conference we had just extended the $15\,\mu$m galaxy source counts an
order of magnitude fainter than any other surveys, exploiting the
lensing magnifications (Hopwood et al., 2010). Since then, 
Hopwood et al., (in prep.) have performed bespoke analysis of the
SPIRE submm data in the field, stacked the submm fluxes of the AKARI
$15\,\mu$m-selected population, and found the $15\,\mu$m population is
responsible for $\sim40-30\%$ of the FIRAS backgrounds at
$250-500\,\mu$m, with the uncertainties dominated by the $5\sigma$
infrared background measurements. 

\begin{figure}[!ht]
\resizebox{1.3\hsize}{!}{\includegraphics{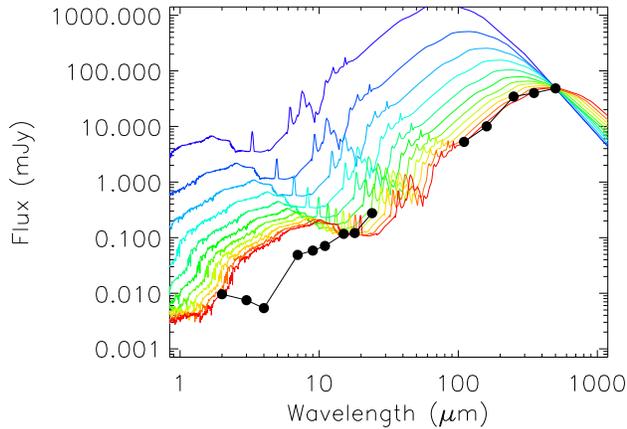}}
\caption{\label{fig:nep} 
M82 local template, normalised to the $500\,\mu$m confusion limit,
from $z=0$ increasing in steps of $\delta z=0.5$ (blue to red). Also
shown are the SPIRE confusion limits, the NEP-Deep AKARI depths, and
the PACS depths assuming all priority 2 scheduled data is
obtained. Note that the AKARI depths are more than sufficient to
detect the submm-selected population.
}
\end{figure}

It should not be altogether surprising that there are such strong
links between the mid-infrared and submm populations. To demonstrate
this, we return finally to the first local galaxy discussed in this
paper, M82. Figure \ref{fig:nep} shows redshifted M82 template SEDs
normalised to the $500\,\mu$m confusion limit compared to the AKARI
NEP-Deep mid-infrared depths. The optical stellar populations will
obviously not be representative of the high-redshift population but
the mid-infrared fluxes should be realistic. The AKARI depths
are clearly easily deep enough to detect the the $500\,\mu$m
population with this SED.

The NEP field has already been observed with Herschel in a
$9\,$deg$^2$ map, with more data still scheduled in priority 2
time. SCUBA-2 and LOFAR data are also imminent, and the field has been
the target of many other multi-wavelength campaigns. In the longer
term, the Euclid mission (launch 2019) will, in addition to its
$20,000$\,deg$^2$ wide survey, devote $40$\,deg$^2$ to deep
cosmological fields. The location or locations of these
$40\,$deg$^2$ are determined partly by the need to cover famous fields
with lots of multi-wavelength legacy data, and partly by the
constraints of the scanning strategy of the mission. The latter
in particular very strongly favours the ecliptic poles,
which are the natural deep field locations for many space
observatory and survey missions. 

\newsection{CONCLUSIONS}
  Preliminary photometry for galaxies in the Herschel Reference
  Survey demonstrates dust masses typically a few $\times10^7M_\odot$.
  These would be unmeasurable without the AKARI all-sky
  survey. However, systematic errors of unknown origin of the order
  $\sim30\%$ are still present in at least some photometric
  measurements; work is ongoing to improve this photometry.
  Preliminary photometry for galaxies in the JCMT Nearby Galaxies
  Survey extends the far-infrared:CO(3-2) luminosity-luminosity
  correlation down five orders of magnitude in homogenously-selected
  samples. 
  The molecular gas depletion timescale for JCMT Nearby Galaxies
  Survey targets is typically \mbox{$\sim3\,$Gyr}, about an order of
  magnitude longer than in high-redshift submm-selected galaxies.
  The AKARI ultra-deep $15\,\mu$m population contributes about a
  third or more of the extragalactic background light at
  $250-500\,\mu$m. 

\acknowledgments{We thank the anonymous referee for helpful comments. SS thanks STFC for financial support under grants
  ST/G002533/1 and ST/J001597/1. The Second AKARI Conference was
  supported by BK21 program to Seoul National University by the
  Ministry of Education, Science and Technology, Center for Evolution
  and Origin of Universe (CEOU) at Seoul National University, National
  Research Foundation Grant No. 2006-341-C00018 to HMLee, Astronomy
  Program, Seoul National University Nagoya University Global COE
  Program: Quest for Fundamental Principles in the Universe, Division
  of Particle and Astrophysical Science, Nagoya University, and
  Institute of Space and Astronautical Science, Japan Aerospace
  Exploration Agency. This research has made use of the NASA/IPAC
  Extragalactic Database (NED) which is operated by the Jet Propulsion
  Laboratory, California Institute of Technology, under contract with
  the National Aeronautics and Space Administration. }

\end{description}

\end{document}